# A novel method to measure the dE/dx resolution of a TPC prototype using 266 nm UV laser

**Yiming Cai**[a,b]**, Yulan Li**[a,b,*]**, Huirong Qi**[c,†]**, Yuanjing Li**[a,b]**, Jian Yang**[a,b]**, Zhiyang Yuan**[c,d]**, Yue Chang**[c]**, Zhi Deng**[a,b]**, Hui Gong**[a,b]**, Jian Zhang**[c]**, Yuanbo Chen**[c] **and Jin Li**[c].

[a] *Department of Engineering Physics, Tsinghua University, Beijing 100084, China*

[b] *Key Laboratory of Particle & Radiation Imaging (Tsinghua University), Ministry of Education Beijing 100084, China*

[c] *Institute of High Energy Physics, Chinese Academy of Science, Beijing 100049, China*

[d] *University of Chinese Academy of Sciences, Beijing 100049, China*

  *E-mail:* `yulanli@mail.tsinghua.edu.cn; qihr@ihep.ac.cn`

ABSTRACT: The specific energy loss (dE/dx) resolution is critical for particle identification and separation in TPC. In connection with the R&D for the TPC at the Circular Electron Positron Collider (CEPC), we find the laser ionisation obeys the Gaussian distribution theoretically and experimentally. Based on this, we develop a novel method using a 266 nm UV laser to measure the dE/dx resolution. Comparing to the traditional methods with particle beams or cosmic rays, the novel method with 266 nm UV laser is more convenient and flexible. We verify the method's feasibility with a TPC prototype and extrapolate the dE/dx resolution to the CEPC TPC. The estimation of the dE/dx resolution is $4.99 \pm 0.16\%$ without the magnetic field.

KEYWORDS: dE/dx resolution; UV lasers; TPC; CEPC.

---

[*] Corresponding author.
[†] Corresponding author.

# Contents

## 1. Introduction

A Time Projection Chamber (TPC) is the main tracking detector in many large particle physics experiments, such as STAR [1], ALICE [2], and ILC [3]. One of the remarkable features of Time Projection Chamber (TPC) is its powerful charged particle identification (PID) capability [4] due to its ability to provide the dE/dx and the momentum of a charged particle simultaneously. The key factor influencing the separation power between two particle species is the dE/dx resolution. An empirical formula describing the dE/dx resolution of present and past large detectors can be derived by fitting the available resolutions as a function of the pad geometry and pressure [5]:

$$\frac{\sigma(dE/dx)}{dE/dx} = 5.5 \times L^{-0.36} \ (\%)$$

with $L = N_{samples} L_{samples} \times \text{pressure (m bar)}$.

There are several traditional methods evaluating the dE/dx resolution. The first utilizes charged particle beams such as electrons, pions, kaons [1][6] and is the most commonly used and robust one. Rare particle beam facilities and high usage expenses are apparent shortcomings of the method. In comparison, the second method with cosmic rays [7] is accessible. Nevertheless, a necessary complicated cosmic ray trigger that simultaneously records cosmic ray energy and a long experimental period are still a barrier. The third method is through simulations [8], which is convenient, and allows to study possible factors influencing the detector dE/dx resolution. However, simulation results do not always agree perfectly with experimental results. There is a drawback in traditional methods that the ionisation probability density distributions (also called straggling function [9]) of charged particles are Landau. We have to do truncation in the calculation of dE/dx resolution which abandons large quantities of events and produce a biased estimator for dE/dx.

UV lasers are low-cost, transportable, and flexible alternatives to particle beams. Moreover, one can conveniently change laser ionisation by adjusting laser energy. Given that lasers can generate the same amount of electron-ion pairs as charged particles, they have the potential to simulate those particles with different momentum under diverse physical requirements in the research of TPC dE/dx resolution. Though UV lasers have been used for distortion calibration in several TPCs [1][2] because of the outstanding advantages, they are not widely applied to energy calibration mainly due to the large fluctuation of laser ionisation. This study proposes a novel method to remove this obstacle by monitoring each laser pulse's energy and correcting corresponding laser ionisation through a power function from our previous research [10]. After such an energy correction, the probability density distribution of laser ionisation becomes Gaussian without having to apply the truncated mean method. The dE/dx resolution can be defined as $\sigma/\mu$, where $\sigma$ and $\mu$ are the standard deviation and the mean of the Gaussian distribution, respectively. An experiment using a TPC with triple Gas Electron Multipliers (GEM) has been conducted to verify the method. The experiment result can further be used to extrapolate to the CEPC TPC, which has an inner and outer radius of 0.3 m and 1.8 m, respectively, and has 220 pad rows along the radial direction [11]. The estimation of CEPC TPC dE/dx resolution is $4.99\pm0.16\%$, fulfilling the requirement of a 2-4 $\sigma$ separation of Κ/π for momentum between 2-20 GeV.

## 2. The method to measure the dE/dx resolution using UV lasers

Nd: Yag lasers (frequency-quadrupled to 266 nm) and nitrogen lasers (337 nm) are the most frequently used lasers in gas detectors. These UV lasers can form ionised tracks by ionising organic impurities in working gases through the two-photon ionisation process [12]. In most experiments, the average ionisation densities of UV lasers are approximately several MIPs [2][13][14][15] to fit the dynamic range of the front-electronics. By comparing positions of reconstructed tracks and pre-defined lasers, one can precisely calculate track distortions. Combined with characteristics of high stability and repeatability, UV lasers are becoming the most effective tool for distortion calibration in gas detectors, especially TPC. However, UV lasers are not usually used in energy-related studies, such as measurements of the dE/dx resolution, due to the large FHWM and Landau-like tail of the measured distribution of the ionisation. In our previous study [10], the measured distribution was fitted with a Landau. A resolution of the laser ionisation, defined as scale parameter divided by the Most Probable Value (MPV) of approximately 25% was found, which is not good enough for energy-related studies. We need to find a way to suppress the laser ionisation fluctuation.

### 2.1 A brief theoretical derivation of the laser ionisation distribution

The laser ionisation probability density distribution has been approximated as a Poisson distribution in [12], which does however conflict with the measured Landau-like tail. Thus, we should first derive the laser ionisation theoretically. The derivation refers to the charged particles ionisation. As mentioned above, the mechanism of laser interaction is the two-photon ionisation [12]. The absorption of photons by impurity molecules is an absolutely random process, which is the same as encounters of charged particles with gas atoms. This process can be characterized by a mean free flight path:

$$\lambda = 1/(N\phi^2\sigma^{(2)})$$

Where N is the density of impurity molecules, $\phi$ is the photon flux and $\sigma^{(2)}$ is the two-photon absorption cross-section. Each absorption happens independently, so the reaction frequency distribution along any length L should follow the Poisson distribution the same as charged particles:

$$P(L/\lambda, k) = (\frac{L/\lambda}{k!})^k e^{(-L/\lambda)}$$

Each absorption of two photons only generates one electron, which differs a lot from the charged particles with complex mechanisms of energy deposition and process of electron-ion pairs generation. Then, the number of ionisation electrons per unit length should also follow the Poisson distribution. In most experiments, the number of electrons from laser ionisation is larger than $100/cm$, in which case the Poisson distribution can be replaced by the Gaussian distribution.

**2.2 Energy correction**

We assume that the discrepancy between the theoretical Gaussian distribution and the measured result is mainly due to the fluctuation of laser energy, which is inevitable for pulsed lasers. As shown in our previous work [10], the relation between the laser energy density and the ionisation density is a power law. The exponent has been extracted by fitting, and is 2.35 for the T2K gas. Therefore, small laser enegy fluctuations may lead to large ionsation fluctuations. As a consequence, laser pulses which energy deviate much from the mean energy causes the long tail. In our experiment, the max energy of the laser before any optical device (initial laser) is around $20$ mJ, but it has to be reduced significantly to around $2$ mJ to ensure the laser ionisation equivalent to several MIPs in TPC, which in turn increases the laser energy fluctuation, and as a consequence, the laser ionsation fluctuation.

As already discussed, the ionisation probability density distribution for a monoenergetic laser is expected to be Gaussian. However, the practical laser energy fluctuates with time, so the measured ionisation spectrum is Landau-like with a large FHWM derived from the superposition of many Gaussian distributions, each corresponding to given energy. If the ionisation of each laser pulse is corrected based on the average laser energy, the superposition of different distributions can approach the case of the monoenergetic laser. This method named energy correction will reduce large FHWM and improve the dE/dx resolution greatly. In Section 3 we will describe how the energy of each laser pulse can be accurately measured.

**3. Experimental setup**

A TPC prototype with triple standard CERN GEMs ($100 \times 100$ mm$^2$) is used to experimentally verify the method. The detector is operated in T2K gas (Ar:CF$_4$:iC$_4$H$_{10}$= 95:3:2) [16] at atmospheric pressure and room temperature. There are 128 readout pads in several rows, and the neighboring rows are staggered with a pitch of half a pad. Each pad is 6 mm long and 1 mm wide. The triple GEM amplification structure, the pad geometry, the working gas and temperature and pressure are the same as for the CEPC TPC. The signals collected by the readout pads are first processed by the front-end electronics integrated with preamplifiers and shapers [17], and then digitized by the data acquisition (DAQ) system [18]. The experiment uses Q-smart 100 pulsed Nd: Yag laser from Quantel Ltd, and the laser wavelength is quadrupled to 266 nm. The optical system consists of a (99/1) partially reflective mirror, two diaphragms, and a beam expander that collimates and attenuates the pulsed lasers. The laser spot size is narrowed to 0.8 mm, and the laser energy density is reduced to approximately $1~2$ μJ/mm$^2$. The laser enters the detector through quartz windows on the detector's sidewalls. Figure 1 shows the layout of the experiment. Further details of the experimental setup, including the detector, electronics, gain calibration, and the optical system can be found in [10].

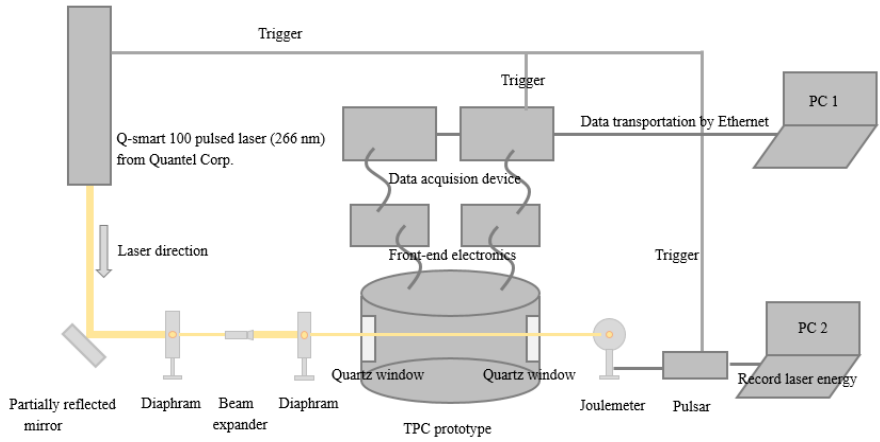

**Figure 1.** The layout of the experiment. The laser is attenuated by a partially reflective mirror and then collimated by two diaphragms. There are two quartz windows with high transmittance introducing the laser into the TPC. The readout pads are on the upper part of the detector connected by the front end electronics. A T-joint BNC and two coaxial cables are used to transmit the trigger signals to the DAQ and Pulsar that accepts the external trigger signal and sends the energy measured by the energy meter to the PC separately.

The trigger is generated by the laser and sent to DAQ for detector signals acquisition, and to a dedicated readout card (Ophir Pulsar) for recording the laser pulse energy. The frequency of the laser is 20 Hz, so the delay in transmission of two trigger signals (in order of ns) does not affect the one-to-one correspondence between the laser pulse energy and the laser ionisation signals.

## 4. Result

Nearly 43000 events have been recorded during 40 minutes for analysis. The detector gain is calibrated by a $Fe^{55}$ source and set to 1300. The gain calibration also indicates that there are several dead channels, which means eight pad rows can be used for subsequent data processing.

### 4.1 The experimental result of the dE/dx resolution

For each laser incidence event, the charges collected from 8 available pad rows are averaged to calculate a dE/dx value. This value is then corrected using the method mentioned in Section 2. It is worth pointing out that the events with energy out of the range [$\sigma+\mu$, $\sigma-\mu$] are filtered out to optimize the calculation of dE/dx value, and this procedure discards about 33% of all events. Figure 2 shows the laser ionisation probability density distribution before and after laser energy correction and events filtering. The improved distribution after laser energy correction and events filtering is actually a Gaussian distribution. The dE/dx resolution can be determined directly by the mean and the standard deviation, and the value is $13.70 \pm 0.07\%$.

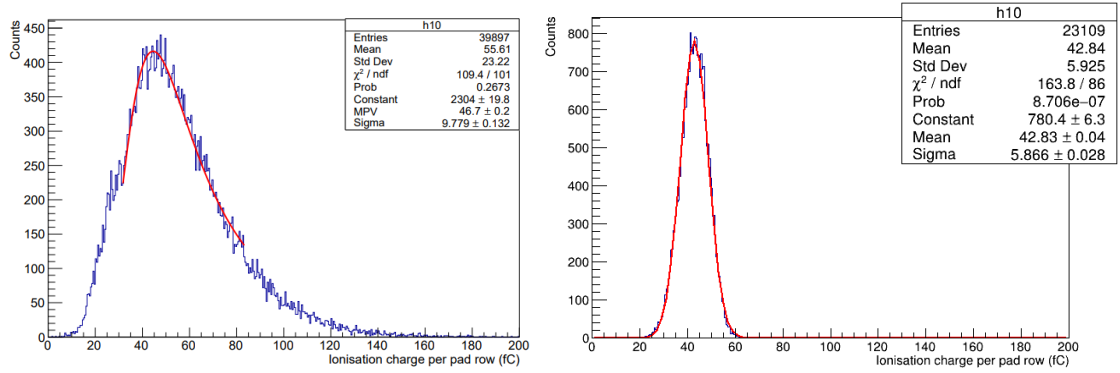

**Figure 2.** The 266 nm laser ionisation probability density distribution (a) original. (b) after laser energy correction and events filtering. The original distribution has large FHWM and Landau-like tail. After laser energy correction and events filtering, the distribution turns to the Gaussian distribution. The mean value stands for the actual average energy loss of the laser, and the dE/dx resolution can be directly defined as the standard deviation devided by the mean.

### 4.2 Extrapolation of the dE/dx resolution to CEPC TPC

The energy correction allows to extrapolate the 8-pad-row experimental result to a system with more pad rows. This is done by combining several events together. As shown in [6], the achieved dE/dx resolution does not depend on the drift length. Twenty-eight events are combined together to calculate the dE/dx resolution, and the estimation result is $4.99 \pm 0.16\%$, as shown in Figure 3.

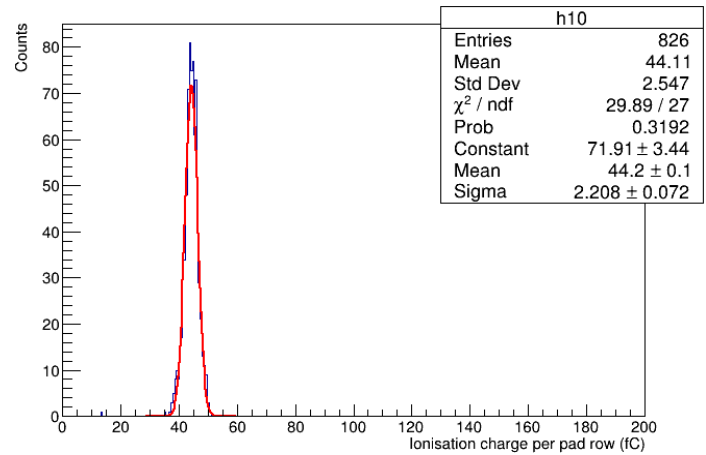

**Figure 3.** The estimation of the dE/dx resolution for the CEPC TPC is $4.99 \pm 0.16\%$. Twenty-eight events (224 pad rows) are combined together to simulate one event in the CEPC TPC (220 pad rows).

We also simulate events with a different number of pad rows. The blue dots in Figure 4 represent the extrapolation results and show the expected strong dependency of the dE/dx resolution on the number of samples (pad rows). A power function with exponent around -0.3 can describe this tendency. The red dot corresponds to the dE/dx resolution measured with the 8-pad-row prototype.

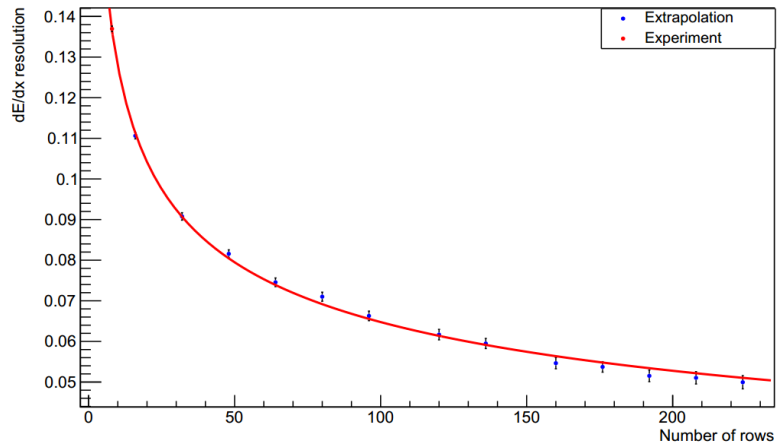

**Figure 4.** The red dot corresponds to the 8-pad-row experiment result. The blue dots show the data from the extrapolation to larger numbers of rows. The points are fitted by a power function (red line). The error bars increase for larger numbers of rows, because more events are combined, and the number of extrapolated events decreases, which increases the statistical error.

## 5. Conclusion

The laser ionisation probability density distribution used to be considered as a Landau distribution similar to charged particles or a Poisson distribution. We find that it is actually a Gaussian distribution, which is experimentally verified using an energy correction method. This discovery reveals that lasers, a perfect tool with many unique advantages, can play a critical role in energy-related studies in gas detectors. We use a TPC prototype to measure the dE/dx resolution and extrapolate it to CEPC TPC. The preliminary result is $4.99 \pm 0.16\%$ without the magnetic field. It shows the validity of the design of the CEPC TPC to some extent.

The next stage of work is to obtain a more accurate dE/dx resolution for CEPC TPC with a larger TPC and magnetic field. Futhermore, we plan to use UV lasers with different energy to verify the UV laser's potential to simulate particles with different momentum.

## Acknowledgments

This work was supported by the National Natural Science Foundation of China under Grant Nos. 11535007 and Grant Nos. 11775242 and the National Key Programme for S&T Research and Development under Grant No. 2016YFA0400400.